\documentclass[12pt]{article}
\usepackage{graphicx}
\usepackage[cp1251]{inputenc}
\begin{document}
\pagestyle{empty} \pagestyle{headings} {\title{Neutrino
oscillations in vortex and twisting magnetic fields}}
\author{O.M.Boyarkin$^a$\thanks{E-mail:oboyarkin@tut.by},
\ I.O.Boyarkina\thanks{E-mail:estel20@mail.ru}\ $^b$\\
$^a$\small{\it{Belarusian State University,}}\\
\small{\it{Dolgobrodskaya Street 23, Minsk, 220070, Belarus}}\\
$^b$\small{\it{The University of Tuscia}}\\
\small{\it{DEIM, 47, Paradise str, Viterbo, Italy}}}
\date{}
\maketitle

\begin{abstract}

The behavior of the neutrino flux in vortex and twisting magnetic
fields is considered within the left-right symmetric model. By way
of illustration of the magnetic fields we discuss the magnetic
fields of the coupled sunspots (CS's) which are the sources of the
future solar flares. It is expected that the neutrinos have such
multipole moments as the charge radius, the magnetic and anapole
moments. The evolution equation in the Schrodinger-like form is
found and all magnetic-induced resonance conversions are analyzed.
It is demonstrated that in the case of the high-energy flares the
sizeable depletion of the $\nu_{eL}$ neutrinos caused by their
resonance absorptions takes place. Possibilities of observations
of this phenomena are investigated at neutrino telescopes whose
work is based on the reaction of the coherent elastic
neutrino-nucleus scattering.
\\[1mm]
PACS number(s): 12.60.Cn, 14.60.Pg, 96.60.Kx, 95.85.Qx, 96.60.Rd.
\end{abstract}
\hspace{1mm}Keys words: Solar flares, flare forecasting, neutrino
oscillations, magnetic and anapole moments, charge neutrino
radius, neutrino telescopes, coherent elastic neutrino-nucleus
scattering, RED-100.

\section{Introduction}

Interaction of neutrinos with external electromagnetic fields is
defined by the multipole moments (MM's) which are caused by the
radiative corrections. The neutrino MM's have been drawing
considerable attention of physicists for many years. However, any
evidences in favor of nonzero neutrino MM's both from laboratory
experiments with ground-based neutrino sources and from
observations of astrophysical neutrino fluxes were absent. It
should be stressed that until recently all that has been measured
was the measurement of the upper bounds on the MM values. The
situation was reversed after the XENON collaboration presented
results of the search for new physics with low-energy electronic
recoil data obtained with the XENON1T detector \cite{EA20}. One of
the possible explanations of their results allows the presence of
a sizable neutrino magnetic moment having the value of the order
of the existing laboratory bounds. It should be stressed that due
to smallness of the MM's the electromagnetic interaction of a
neutrino becomes essential in the case of intensive fields only.
The examples of such fields are the Sun's magnetic fields. In that
case of special interest are the magnetic fields of the sunspots
being the sources of the future solar flares (SF's). The energy
generated during the SF could be as large as $10^{28}-10^{33}$
erg. It is believed that the magnetic field is the main energy
source of the SF's \cite{SI81,KST11}. During the periods of the
high activity of the Sun, the magnetic flux $\sim10^{24}\
\mbox{G}\cdot\mbox{cm}^2$ \cite{DJ81} is thrown up from the solar
interior and accumulates within the sunspots. In so doing, big
sunspots of opposite polarity could be paired forming, the
so-called, coupled sunspots (CS's). Then the process of magnetic
energy storage begins. The length of the initial SF stage is
extended from several to dozens of hours. The data concerning
centimeter radiation above a spot is indicative of the gas heating
up to the temperatures of a coronal order leading to a high value
of solar plasma conductivity. Therefore, the longitudinal electric
current $J_z$ might be large enough in a region above sunspots. In
Ref. \cite{KDA15} it was shown that when the magnetic field of
newly emerged sunspot $B_{cs}$ takes the value 2000 G, $J_z$ can
reach $(0.7-4)\times10^{12}$ A. Therefore, for the CS's which
magnetic fields could increase up to the values of $10^6$ G, $J_z$
could be $10^{14}$ A and above. For the big CS's with
$R_s\simeq10^8\ \mbox{cm}$ the electric current density could be
as large as $10^{-1}\ \mbox{A}/\mbox{cm}^2$. The more powerful the
SF is, the greater the magnetic field strength of the CS's will
be. For example, in the case of the super-SF's \cite{ML17}, which
energy could be of order of $10^{36}$ erg, $B_{cs}$ may reach the
values of $10^8$ G.

It is clear that the high-power SF's are very destructive when
they are focussed on the Earth as it was in 1859 (the Carrington
Flare \cite{RAC}). It might be worth pointing out that the flare
events are also at work in other Sun-like stars (first-generation
stars). Flares at these stars are also dangerous for crew members
of interplanetary spaceships. Consequently, for our increasingly
technologically dependent society the SF forecasting has great
practical value. Moreover, the study of the SF's helps to
elucidate the structure and evolution of the Universe as well.
There are special cosmic projects \cite{AOB} which are focussed on
investigation of the flares happening at the Sun-like stars. For
example, by now the Kepler mission \cite{SC14} surveying the
$\sim10^5$ stars has accumulated a great deal of data concerning
the large flares with energies of order of $10^{33}$ erg.

For the most part, the methods of the SF forecasting are based on
measurements of the magnetic fields of the active regions in the
solar photosphere being made with $\gamma$-telescopes observing
the Sun continuously (see, for example \cite{RLM}). Worthy of
mention is the recent breakthrough in the SF predictions made by a
team of Japanese physicists \cite{KK20}. They presented, the
so-called kappa-scheme, a physics-based model forecasting the
large SF's through a critical condition of magnetohydrodynamic
instability, triggered by magnetic reconnection. The group tested
the method using observations of the Sun from 2008 to 2019. In
most cases, the method correctly identifies which regions will
produce large SF's within the next 20 hours, The method also
provides the exact location where each SF will begin and limits on
how powerful it will be.

This paper is a continuation of the works \cite{BB200,BGAL,BGAL1}
in which the correlation between the SF's and behavior of the
neutrino beam was discussed. It is clear that when the electron
neutrinos beam passing through the magnetic field of the CS's will
change its composition and we could detect this changing, then the
problem of the SF's forecasting with a help of neutrino telescopes
will be resolved. In the three flavor basis this problem was
discussed in Ref.\cite{BGAL1} for the Dirac neutrinos. However in
that work it was assumed that the nondiagonal neutrino anapole
moments are equal to zero and the neutrino charge radios was
ignored. Contrary to that work our consideration carries more
general character, namely, we take into account all multipole
moments describing the neutrino interaction with magnetic fields
and do not make any constraints on the anapole moments. Our
analysis will be fulfilled both for the Majorana and Dirac
neutrinos. In the next Section we discus the general form of the
electromagnetic interaction of Majorana and Dirac neutrinos and
address the phenomenology of the neutrino multipole moments in
laboratory experiments. Over the course of the work we assume that
the multipole moments have the values being close to the
experimental limits. In the third Section we obtain the evolution
equation for the neutrino beams in two flavor approximation and
find all the possible resonance conversions. In the forth Section
we do the same working in three flavor basis. Finally, in Section
5, some conclusions are drawn.

\section{Electromagnetic neutrino properties}

In the one-photon approximation, the effective interaction
Hamiltonian satisfying the demands both of the  Lorentz and of the
electromagnetic gauge invariance is determined by the following
expression \cite{Ni82,Ka82}
$${\cal{H}}^{(\nu)}_{em}(x)=
\sum_{i,f}\overline{\nu}_i(x)
\{i\sigma_{\mu\lambda}q^{\lambda}[F^{if}_M(q^2)+iF^{if}_E(q^2)\gamma_5]+
(\gamma_{\mu}-q_{\mu}q^{\lambda}
\gamma_{\lambda}/q^2)[F^{if}_Q(q^2)+$$
$$+F^{if}_A(q^2)q^2\gamma_5]\}\nu_f(x)A^{\mu}(x),\eqno(1)$$
where $q_{\mu}=p_{\mu}^{\prime}-p_{\mu}$ is the transferred
4-momentum, while $F^{if}_Q, F^{if}_M, F^{if}_E,$ and $F^{if}_A$
are the real charge, dipole magnetic, dipole electric, and anapole
neutrino form factors. The form-factors with $i=f$ ($i\neq f$) are
named "diagonal" ("off-diagonal" or "transition") ones. In the
static limit ($q^2=0$), $F^{if}_M(q^2)$, $F^{if}_E(q^2)$ and
$F^{if}_A(q^2)$ determine the dipole magnetic, dipole electric and
anapole moments, respectively. The second term in the expansion of
the $F_Q^{if}(q^2)$ in series of powers of $q^2$ is connected with
the neutrino charge radius
$$<r^2_{if}>=6{dF^{if}_Q(q^2)\over dq^2}\Bigg|_{q^2=0}.\eqno(2)$$
In what follows, amongst the neutrino electromagnetic
characteristics, we shall be interested in the dipole magnetic
moments (DMM), the anapole moments (AM) and the charge radii
(NCR).

For the first time the behavior of the neutrino endowed by the DMM
in the external magnetic field was discussed in Ref. \cite{OVV86}.
Since then many works appeared in which the problems of the solar
neutrinos were investigated with inclusion of the DMM
\cite{CS90,AB93,ekh93, XS93} (see, for Review \cite{GA15}). It
should be recorded that examination of the effects produced by the
neutrino DMM's could help to find out the neutrino nature (Dirac
or Majorana). The Dirac neutrinos may have both the diagonal and
off-diagonal DMM's while the Majorana neutrinos could possess only
the off-diagonal DMM's with a property
$\mu_{ll^{\prime}}=-\mu_{l^{\prime}l}$.

The exhibiting of neutrino DMM's are being searched in the
reactors (MUNU, TEXONO and GEMMA) \cite{ZD05,HTW07,AGB12},in the
accelerators (LSND) \cite{LBA01,RS01}, and in the solar
(Super-Kamiokande and Borexino) \cite{CA08,DWL04} experiments. The
current best sensitivity limits on the diagonal DMM's obtained in
laboratory measurements are as follows
$$\mu_{ee}^{exp}\leq2.9\times10^{-11}\mu_B, \qquad90\%
\ C.L.\qquad [\mbox{GEMMA}] \ \cite{AGB12},$$
$$\mu_{\mu\mu}^{exp}\leq6.8\times10^{-10}\mu_B, \qquad90\%\
C.L.\qquad [\mbox{LSND}]\ \cite{LBA01}.$$ For the $\tau$-neutrino,
the limits on $\mu_{\tau\tau}$ are less limitative (see, for
example \cite{tauB}), and the current upper bound on that is
$3.9\times10^{-7}\mu_B$.

Astrophysical and cosmological arguments are more limitative. For
example, in Ref.\cite{RB88} it was demonstrated that the absence
of high-energy events in the SN1987A neutrino signal leads to the
inequality $\mu_{\nu_e\nu_e}\leq10^{-12}\mu_B$ at 90\%\ C.L.
Cooling rates of red giants \cite{GG99} results a comparable limit
$\mu_{\nu_e\nu_e}\leq3\times10^{-12}\mu_B$ at 90\%\ C.L., whereas
analysis of cooling rates of white dwarfs \cite{SIB94} puts a
bound of $\mu_{\nu_e\nu_e}\leq10^{-11}\mu_B$ at 90\%\ C.L. It
should be stressed that what is measured in real experiments is
the effective DMM $\mu^{exp}_{\nu_l\nu_l}$ whose value is a rather
composite function of the transition magnetic moments. Moreover,
the dipole electric transition moments, if these quantities do not
vanish, could give the contribution to $\mu^{exp}_{\nu_l\nu_l}$ as
well. We emphasize that the reliable bounds on transit DMM's could
be obtained only from detailed studying of the processes with the
partial lepton flavor violation. At present, in the Majorana
neutrino case the global fit of the reactor and solar neutrino
data gave the result \cite{WG03}
$$\mu_{12}, \mu_{13}, \mu_{23}\leq1.8\times10^{-10}\mu_B.\eqno(3)$$

Even though a neutrino has the electric charge being equal to
zero, the neutrino could possesses superposition of two charge
distributions of opposite signs, which is featured by an electric
form factor. Then the second term in the expansion of this form
factor in series of powers of $q^2$ is connected with the NCR. The
NCR influences the processes of the neutrino scattering on charged
particles. The limits on the NCR's could be received from the
studying the elastic neutrino-electron scattering. For example,
investigation of this process at the TEXONO experiment results in
the following bounds on the NCR \cite{TSK15}
$$-2.1\times10^{-32}\ \mbox{cm}^2\leq(<r_{\nu_e}^2>)\leq
3.3\times10^{-32}\ \mbox{cm}^2.\eqno(4)$$ Investigation of
coherent elastic neutrino-nucleus scattering at the TEXONO
(\cite{MD10}), LSND (\cite{LB01}) and BNL-E734 (\cite{LA90})
experiments allowed to obtain the bounds on the diagonal NCR's
$$-4.2\times10^{-32}\ \mbox{cm}^2\leq(<r^2_{\nu_e}>)\leq6.6\times10^{-32}\
\mbox{cm}^2,\qquad[\mbox{TEXONO}]$$
$$-5.94\times10^{-32}\ \mbox{cm}^2\leq(<r^2_{\nu_e}>)\leq
8.28\times10^{-32}\ \mbox{cm}^2, \qquad[\mbox{LSND}]$$
$$-5.7\times10^{-32}\ \mbox{cm}^2\leq(<r^2_{\nu_{\mu}}>)\leq
1.1\times10^{-32}\ \mbox{cm}^2, \qquad [\mbox{BNL-E734}].$$ In its
turn the bounds on the transition NCR's
$$\left.\begin{array}{ll}|<r^2_{\nu_e\nu_{\mu}}>|\leq28\times10^{-32}
\mbox{cm}^2,\qquad|<r^2_{\nu_e\nu_{\tau}}>|\leq30\times10^{-32}
\mbox{cm}^2,\\[2mm]
\hspace{30mm}|<r^2_{\nu_{\mu}\nu_{\tau}}>|\leq35\times10^{-32}\
\mbox{cm}^2.
\end{array}\right\}\eqno(5)$$ were gotten from analysis of the
COHERENT data on CENNS \cite{MC20}.

The NCR affects both on astrophysics and on cosmology. For
example, in the case when the Dirac neutrino has the charge
radios, in the $e^+e^-$ annihilations the right-handed
neutrino-antineutrino pairs could be produced. This process could
influence primordial Big-Bang Nucleosynthesis and the energy
release of a core-collapse supernova.

The AM of $1/2$-spin Dirac particle was introduced in the work
\cite{Y57} for a $T$-invariant interaction which does not conserve
$P$-parity and $C$-parity, individually. Later in order to
describe this kind of interaction a more general characteristic,
the toroid dipole moment (TDM) \cite{VM74}, was entered. It was
shown that the TDM is a general case of the AM and at the
mass-shell of the viewed particle the both moments coincide. The
neutrino toroid interaction are manifested in scattering of the
neutrinos with charged particles. In so doing, the interaction
saves the neutrino helicity and gives an extra contribution, as a
part of the radiative corrections. In this regards, the AM is
similar to the NCR. Both quantities preserve the helicity in
coherent neutrino collisions, but have various nature. They define
the axial-vector (AM) and the vector (NCR) contact interactions
with an external electromagnetic field, respectively. From the
viewpoint of determining the NCR and AM the low-energy scattering
processes are of special interest (see, for example, Refs.
\cite{JL85,RC91}).

The both neutrino interactions may have very interesting
consequences in various media. The possible role of the AM in
studying the neutrino oscillations was first specified in Refs.
\cite{VD98}). A point that should be also mentioned is Ref.
\cite{OMDR} where the behavior of neutrinos endowed with the AM in
a vortex magnetic field was considered upon discussing the
correlation between the electron neutrino flux and the solar flare
events.

Since phenomenology of the AM is analogous to that of the NCR, the
linkage between these quantities must exist. In the SM for a
zero-mass neutrino, the value of the AM $a_{\nu}$ is connected
with the NCR through the simple relation (see, for example,
\cite{AR00})
$$a^{\prime}_{\nu}={1\over6}<r_{\nu}^2>\eqno(6)$$
(the dimensionality of the AM in CGS system is
"$\mbox{length}^2\times\mbox{charge}$", that is to say,
$a_{\nu}=ea^{\prime}_{\nu}$ \cite{Y57}). However even in the SM
with the massive neutrinos this relationship is violated
\cite{MSD04}. It breaks down in the case of the SM extensions as
well \cite{MSD04}. Mention should be also made of the relation
$$a_{\nu_e}\simeq
e{\sqrt{2}G_F\over\pi^2}=8.5\times10^{-13}\mu_B\lambda_e,\eqno(7)$$
($\lambda_e$ is an electron Compton wavelength), which is widely
met in literature (see, for example, \cite{VM98}). It appears to
be very convenient for comparison of interactions caused by
nonzero values of the AM and the DMM with external magnetic field.

\section{Two-flavor approximation}

In the SM the DMM of the neutrino appears to be proportional to
the neutrino mass \cite{KF80}
$$\mu_{\nu}=10^{-19}\mu_B
\Big({m_{\nu}\over\mbox{eV}}\Big),\eqno(8)$$ and, as a result,
cannot bring to any observable effects in real fields. Therefore,
when one employs the values of the neutrino DMM's which are close
to the upper experimental bounds, then one should fall outside the
scope of the SM. To obtain the large value of the neutrino DMM the
SM extension must involve the right-handed charged currents and/or
singly-charged Higgs bosons. As an example of such SM extension we
shall utilize the left-right symmetric model (LRM) based on the
$SU(2)_R\times SU(2)_L\times U(1)_{B-L}$ gauge group \cite{ICP74,
RNM75,GSRN}. In the LRM the Higgs sector content defines the
neutrino nature. If it contains the bi-doublet $\Phi(1/2,1/2,0)$
and two triplets $\Delta_L(1,0,2)$, $\Delta_R(0,1,2)$ \cite{RN81}
(in brackets the values of $S^W_L, S^W_R$ and $B-L$ are given,
$S^W_L$ ($S^W_R$) is the weak left (right) isospin while $B$ and
$L$ are the baryon and lepton numbers, respectively), then the
neutrino represents a Majorana particle. For the neutrino to have
a Dirac nature, the Higgs sector must consist of the bi-doublet
$\Phi(1/2,1/2,0)$ and two doublets $\chi_L(1/2,0,1)$,
$\chi_R(0,1/2,1)$ \cite{RMS77}.

In the LRM the contributions to the neutrino DMM are caused both
by gauge bosons $W^{\pm}$, $W^{\prime\pm}$ and by singly charged
Higgs bosons $h^{(\pm)}$, $\tilde{\delta}^{(\pm)}$
\cite{OMB13,OMB14,OMB19}. Since the masses of $W^{\prime\pm}$ and
$h^{(\pm)}$ are at the TeV scale \cite{MT19}, then one may neglect
their contributions to the neutrino DMM. Alternatively, the
${\tilde{\delta}}^{(\pm)}$ boson does not interact with quarks,
and as a consequence, the more firm data for obtaining the bounds
on the $m_{\tilde{\delta}}$ come from investigation of the
electroweak processes. For example, data of LEP experiments
(ALEPH, DELPHI, L3, and OPAL) yield the bound $m_{H^+}>80$ GeV
\cite{MT19}. The interaction between neutrino and
${\tilde{\delta}}^{(\pm)}$ boson is determined by the Lagrangian
\cite{OMB19}
$${\cal{L}}_{\tilde{\delta}}={f_{ll^{\prime}}\over\sqrt{2}}\overline{l}^c(x)
(1-\gamma_5)\nu_{l^{\prime}}(x)\tilde{\delta}^+(x),\eqno(9)$$
where $f_{ll^{\prime}}$ are triplet Yukawa coupling constants
(TYCC), $l,l^{\prime}=e,\mu,\tau$ and the upper index $c$ means
the charge conjugation operation. This interaction effects
changing of the matter potential on the value
$$V^{\tilde{\delta}}_{ll^{\prime}}=-{f_{el}f_{el^{\prime}}\over
m_{\tilde{\delta}}}n_e,\eqno(10)$$ ($n_e$ is an electron density
of a matter under consideration). The analysis shows that in the
Sun conditions $V^{\tilde{\delta}}_{ll^{\prime}}$ could change the
SM prediction on the value of $\mbox{few}\times10\%$ \cite{BB200}.
In what follows, for the sake of simplicity, we shall assume that
only the diagonal TYCC are different from zero.

As for the magnetic field  of the Sun, we reason that it is
nonhomogeneous and vortex. We also assume that it exhibits the
geometrical phase $\Phi(z)$ (twisting)
$$B_x\pm iB_y = B_{\bot}e^{\pm i\Phi(z)}.\eqno(11)$$
We notice that both  for the Sun and for the Sun-like stars the
reason of twisting is differential rotation rates of their
components and the global convection of the plasma fluid. It
should be noted that configurations of the solar magnetic field
implying twisting nature are already being discussed in the
astrophysical literature for a long time (see, for example
\cite{NY71}). In Ref. \cite{VW90} the phase $\Phi$ was introduced
for the solar neutrino description for the first time.
Subsequently, in Ref. \cite{SM91} an account of this phase was
demonstrated. It should be remarked the works
\cite{AS91,ST91,ekh93} which were devoted to the effects on
neutrino behavior in the twisting magnetic fields. For example, in
Ref. \cite{ekh93} the neutrino beam traveling in the twisting
magnetic fields of the solar convective zone was considered and
some new effects (changing the energy level scheme, changing the
resonances location, appearing the new resonances, merging the
resonances and so on) were predicted. Assuming that the magnitude
of the twist frequency $\dot{\Phi}$ is characterized by the
curvature radius $r_0$ of the magnetic field lines,
$\dot{\Phi}\sim1/r_0$, while $r_0$ has the order of $10\%$ of the
solar radius, the authors came to the following conclusion. To
ensure that these new effects will be observed the value of
$\dot{\Phi}$ in the convective zone should have the order of
$10^{-15}$ eV.

Since we are going to take into account the interaction of the
neutrinos with the electromagnetic fields, the neutrino system
under consideration must include both the left-handed and
right-handed neutrinos. By virtue of the fact that the
right-handed Majorana neutrinos are not sterile and interact as
the right-handed Dirac antineutrinos, we shall denote them as
$\overline{\nu}_{lR}$. In order to stress the sterility of
right-handed Dirac neutrinos we shall use for them the notation
$\nu_{lR}$. So, in two-flavor approximation the Majorana neutrino
system will be described by the function
$\psi^{MT}=(\nu_{eL},\nu_{\kappa L},{\overline\nu}_{eR},
{\overline\nu}_{\kappa R})$ while in the Dirac neutrino case we
deal with the function $\psi^{DT}=(\nu_{eL},\nu_{\kappa
L},\nu_{eR},\nu_{\kappa R})$. In what follows to be specific, we
shall reason $\kappa=\mu$.

In order to facilitate the evolution equation for the solar
neutrinos we pass into the reference frame rotating with the same
angular velocity as the transverse magnetic field. The matrix of
the transition to the new reference frame has the form
$$S=\left(\matrix{e^{i\Phi/2}&           0 & 0 & 0\cr
0 & e^{i\Phi/2}& 0 &0\cr 0 &0& e^{-i\Phi/2}& 0\cr 0&0&0&
e^{-i\Phi/2}}\right).\eqno(12)$$ In this reference frame for the
Majorana neutrino the evolution equation will look like
$$i{d\over dz}\left(\matrix{\nu_{eL}\cr\nu_{\mu L}\cr\overline{\nu}_{eR}
\cr\overline{\nu}_{\mu
R}}\right)=\Big({\cal{H}}_0^M+{\cal{H}}^M_{int}\Big)
\left(\matrix{\nu_{eL}\cr\nu_{\mu
L}\cr\overline{\nu}_{eR}\cr\overline{\nu}_{\mu
R}}\right),\eqno(13)$$ where
$${\cal{H}}_0^M=\left(\matrix{-\Delta^{12}c_{2\theta}&\Delta^{12}s_{2\theta}&
0&0\cr\Delta^{12}s_{2\theta}&\Delta^{12}c_{2\theta}&0&0\cr
0&0&-\Delta^{12}c_{2\theta}&\Delta^{12}s_{2\theta}\cr
0&0&\Delta^{12}s_{2\theta}&\Delta^{12}c_{2\theta}\cr} \right)$$
$${\cal{H}}_{int}^M=\left(\matrix{V_{eL}^{\prime}+{\cal{A}}^L_{ee}-\dot{\Phi}/2
&{\cal{A}}^L_{e\mu}&0&\mu_{e\mu}B_{\perp}\cr {\cal{A}}^L_{\mu
e}&V_{\mu L}+
{\cal{A}}^L_{\mu\mu}-\dot{\Phi}/2&-\mu_{e\mu}B_{\perp}&0\cr
0&-\mu_{e\mu}B_{\perp}&-V^{\prime}_{eL}+{\cal{A}}^R_{ee}+\dot{\Phi}/2&
{\cal{A}}^R_{e\mu}\cr \mu_{e\mu}B_{\perp}&0&{\cal{A}}^R_{\mu
e}&-V_{\mu
L}+{\cal{A}}^R_{\mu\mu}+\dot{\Phi}/2\cr}\right),\eqno(14)$$ the
free Hamiltonian ${\cal{H}}_0^M$ describes oscillations in vacuum,
while the interaction Hamiltonian ${\cal{H}}^M_{int}$ covers
interaction with medium, $V^{\prime}_{eL}$ ($V_{\mu L}$) is a
matter potential describing interaction of the $\nu_{eL}$
($\nu_{\mu L}$) neutrinos with dense matter,
$$V_{eL}^{\prime}=V_{eL}+V^{\tilde{\delta}}_{ee},\qquad
V_{eL}=\sqrt{2}G_F(n_e-n_n/2),\qquad V_{\mu L}=V_{\tau
L}=-\sqrt{2}G_Fn_n/2,$$
$$\Delta^{12}=\frac{m^2_1-m^2_2} {4E},\qquad{\cal{A}}^L_{ll^{\prime}}=
\Big\{e[1-\delta_{ll^{\prime}}]{<r_{\nu_{lL}\nu_{l^{\prime}L}}^2>\over6}+
a_{\nu_{lL}\nu_{l^{\prime}L}}\Big\}[\mbox{rot}\ {\bf{H}}(z)]_z,$$
$$\cos{2\theta}=c_{2\theta},\ \ \sin{2\theta}=s_{2\theta},
\qquad{\cal{A}}^R_{ll^{\prime}}=\Big\{e[1-\delta_{ll^{\prime}}]
{<r_{\overline{\nu}_{lR}\overline{\nu}_{l^{\prime}R}}^2>
\over6}-a_{\overline{\nu}_{lR}\overline{\nu}_{
l^{\prime}R}}\Big\}[\mbox{rot}\ {\bf{H}}(z)]_z,$$
$$m_1=m_e\cos\theta-m_{\mu}\sin\theta,\qquad
m_{2}=-m_e\sin\theta+m_{\mu}\cos\theta,$$ $\theta$ is a neutrino
mixing angle in vacuum, $m_1$ and $m_2$ are mass eigenstates,
$\dot{\Phi}$ is the twisting frequency, and $n_n$ is neutron
density. When writing ${\cal{H}}_{int}^M$ we have taken into
account that the toroid interaction is different from zero in the
presence of the inhomogeneous vortex magnetic field. In a concrete
experimental situation this field could be realized owing to
Maxwell's equations as the displacement and conduction currents.
We can consider the situation with the solar flares (SF's) as an
example. The commonly accepted model of this solar phenomenon is
the magnetic reconnection model \cite{KST11}. According to this
model, a variable electric field induced by magnetic field
variations of the coupled sunspots (CS's) appears at the SF
initial phase. This field causes the conduction current which
takes the form of a current layer directed along limiting strength
line being common for the CS's. So, in this case the neutrinos are
influenced by both the displacement current and the conduction
current.

For the Dirac neutrino flux traveling through the solar medium we
have
$${\cal{H}}^D_0={\cal{H}}^M_0,\qquad
{\cal{H}}^D_{int}=\left(\matrix{V_{eL}+{\cal{A}}^{DL}_{ee}
-\dot{\Phi}/2 &{\cal{A}}^{DL}_{e\mu} & \mu_{ee}B_{\perp}
&\mu_{e\mu}B_{\perp}\cr{\cal{A}}^{DL}_{\mu e}&V_{\mu L}+
{\cal{A}}^{DL}_{\mu\mu}-\dot{\Phi}/2&\mu_{e\mu}B_{\perp}&
\mu_{\mu\mu}B_{\perp}\cr \mu_{ee}B_{\perp}
&\mu_{e\mu}B_{\perp}&\dot{\Phi}/2&0\cr
\mu_{e\mu}B_{\perp}&\mu_{\mu\mu}B_{\perp}&0&
\dot{\Phi}/2\cr}\right),\eqno(15)$$ where
$${\cal{A}}^{DL}_{ll^{\prime}}=\Big\{{e<r_{\nu_{lL}\nu_{l^{\prime}L}}^2>
\over6}+a_{\nu_{lL}\nu_{l^{\prime}L}}\Big\}[\mbox{rot}\
{\bf{H}}(z)]_z,$$ and we have neglected contribution to the matter
potential coming from the singly charged Higgs boson.

Our next task is to investigate the resonance conversions of the
neutrino beam which travels in the region of the CS's being the
source of the solar flares. Remember, that for the resonance
conversion to take place, there is a need to comply with the
following requirements: (i) the resonance condition must be
fulfilled; (ii) the resonance width must be nonzero; (iii) the
neutrino beam must pass a distance comparable with the oscillation
length.

In order to find the exact expressions for the resonance
conversion probabilities we must choose the definite coordinate
functions for the description of quantities $V_{eL}$, $B_{\perp}$,
$\dot{\Phi}$ and solve Eq.(13). Then, with the help of the found
functions $\nu_l(z)$, we could determine all resonance conversion
probabilities. Of course we shall be dealing with numerical
solution and, as a result, the physical implications will be far
from transparent. Moreover, in the most general case some of the
resonance transitions may be forbidden. Therefore, first we must
establish which of these transitions are allowed and which are
forbidden. Further we shall follow generally accepted scheme (see,
for example, \cite{XS93,ekh1993}), namely, we shall believe that
all resonance regions are well separated what allows us to
consider them as independent ones. As far as the twisting is
concerned, amongst existing the twisting models (see, for example
\cite{TKUB94}) we choose the simple model proposed in Ref.
\cite{CA92}
$$\Phi(z)={\alpha\over L_{mf}}z,\eqno(16)$$
where $\alpha$ is a constant and $L_{mf}$ is a distance on which
the magnetic field exists.

We start with resonant conversions of the electron neutrinos in
the Majorana neutrino case. Here the $\nu_{eL}$ may exhibit two
resonance conversions. The $\nu_{eL}\to\nu_{\mu L}$
(Micheev-Smirnov-Wolfenstein
--- MSW \cite{wol78,mikh85}) resonance is the first one. The
corresponding resonance condition, the transition width and the
oscillation length are defined by the expressions
$$\Sigma_{\nu_{eL}\nu_{\mu L}}=
-2\Delta^{12}c_{2\theta}+V^{\prime}_{eL}-V_{\mu L}+
{\cal{A}}^L_{ee}-{\cal{A}}^L_{\mu\mu}=0,\eqno(17)$$
$$\Gamma_{\nu_{eL}\nu_{\mu L}}\simeq
{\sqrt{2}(\Delta^{12}s_{2\theta}+ {\cal{A}}^L_{e\mu})\over
G_F},\eqno(18)$$
$$L_{\nu_{eL}\nu_{\mu L}}={2\pi\over\sqrt{\Sigma_{\nu_{eL}\nu_{\mu L}}^2+
(\Delta^{12}s_{2\theta}+ {\cal{A}}^L_{e\mu})^2}}.\eqno(19)$$ From
Eqs.(18) and (19) it follows that the oscillation length achieves
maximum value at the resonance and the relation
$$\Gamma_{\nu_{eL}\nu_{\mu L}}={2\sqrt{2}\pi\over G_F
[L_{\nu_{eL}\nu_{\mu L}}]_{max}}\eqno(20)$$ takes place. With a
help of the relations (17)-(19) one could obtain the probability
of the $\nu_{eL}\to\nu_{\mu L}$ resonance transition. In the most
simple case, when the neutrino system consists only from
$\nu_{eL}$ and $\nu_{\mu L}$ while the Hamiltonian is not a
distance function, this quantity is defined by the expression
$$P_{\nu_{eL}\nu_{\mu L}}(z)=\sin^22\theta_m\sin^2\Bigg({z\over
L_{\nu_{eL}\nu_{\mu L}}}\Bigg),\eqno(21
)$$ where
$$\sin^22\theta_m=
{(\Delta^{12}s_{2\theta}+{\cal{A}}^L_{e\mu})^2\over{\Sigma_{\nu_{eL}\nu_{\mu
L}}^2+({\Delta^{12}s_{2\theta}+
{\cal{A}}^L_{e\mu}})^2}}\eqno(22),$$ and $\theta_m$ is a mixing
angle in a medium. In order to include contributions from the
lowest-energy but most numerous $pp$-neutrino flux, we put
$E_{\nu}=0.4$ MeV. Next, taking into account $$ \Delta
m^2_{12}=7.37\times10^{-5}\ \mbox{eV}^2,\qquad\
\sin^2\theta=\sin^2\theta_{12}=0.297\eqno(23)$$ we get
$2\Delta^{12}c_{2\theta}\simeq8\times10^{-11}\ \mbox{eV}$. Then
the neutrino flux passing through the region of this resonance
must be reduced by about a factor of two as it was verified by
experiments \cite{Altman}. Since the maximum value of the
oscillation length has the order of $\sim3.5\times10^7$ cm, this
resonance transition is fulfilled before the convective zone.
Consequently, it has no bearing on the SF's which take place in
the solar atmosphere. To put it another way, in the case of the
MSW resonance the quantities ${\cal{A}}^L_{ee}$,
${\cal{A}}^L_{\mu\mu}$ and ${\cal{A}}^L_{e\mu}$ do not play any
role.

Further we shall consider the $\nu_{eL}\to\overline{\nu}_{\mu R}$
resonance. The relations being pertinent to this resonance are as
follows
$$\Sigma_{\nu_{eL}\overline{\nu}_{\mu R}}=-2\Delta^{12}c_{2\theta}
+V^{\prime}_{eL}+V_{\mu L}+{\cal{A}}_{ee}^L-{\cal{A}}_{\mu\mu}^R
-\dot{\Phi}=0\eqno(24)$$
$$\Gamma_{\nu_{eL}\overline{\nu}_{\mu R}}\simeq
{\sqrt{2}(\mu_{e\mu}B_{\perp})\over G_F},\eqno(25)$$
$$L_{\nu_{eL}\overline{\nu}_{\mu R}}\simeq
{2\pi\over\sqrt{\Sigma_{\nu_{eL}\overline{\nu}_{\mu R}}^2+
(\mu_{e\mu}B_{\perp})^2}}.\eqno(26)$$ In the solar atmosphere, the
terms $V^{\prime}_{eL}$ and $V_{\mu L}$ in Eq. (24) are more less
than $\Delta^{12}c_{2\theta}$ and do not play any part.
Analogously the quantity
$(a_{\nu_{eL}\nu_{eL}}+a_{\overline{\nu}_{\mu
R}\overline{\nu}_{\mu R}})[\mbox{rot}\ {\bf{H}}(z)]_z$ appears to
be also small compared with $\Delta^{12}c_{2\theta}$. For example,
in the best case, when the currents producing the inhomogeneous
vortex magnetic field reach the values of $10^{-1}\
\mbox{A}/\mbox{cm}^2$, for the CS's the quantity
$(a_{\nu_{eL}\nu_{eL}}+a_{\overline{\nu}_{\mu
R}\overline{\nu}_{\mu R}})[\mbox{rot}\ {\bf{H}}(z)]_z$  has the
order of $10^{-30}$ eV. Therefore, the resonance
$\nu_{eL}\to{\overline\nu}_{\mu R}$ may occur only at the cost of
magnetic field twisting, that is, when the relation
$$2\Delta^{12}c_{2\theta}+\dot{\Phi}\simeq0.\eqno(27)$$
will be fulfilled.

We can conveniently eliminate the MSW resonance from consideration
assuming that the electron neutrino beam has endured the MSW
resonance before it enters the magnetic field of the CS's. To put
it another way, in what follows we shall deal with the beam which
has already been weakened at the cost of the MSW resonance.

The oscillations picture examined above will be incomplete, if we
don't take into consideration the oscillation transitions of the
$\nu_{\mu L}$ neutrinos, which were produced in the convective
zone due to the MSW resonance. In the magnetic field of the CS's
they could undergo one more resonance conversion, namely, the
$\nu_{\mu L}\to\overline{\nu}_{eR}$ resonance. The resonance
condition and the maximal value of the oscillation length for this
resonance are determined by the expressions
$$\Sigma_{\nu_{\mu L}\overline{\nu}_{eR}}=2\Delta^{12}c_{2\theta}
+V^{\prime}_{eL}+V_{\mu L}+{\cal{A}}_{\mu\mu}^L-{\cal{A}}_{ee}^R
-\dot{\Phi}=0\eqno(31)$$
$$(L_{\nu_{\mu L}\overline{\nu}_{eR}})_{max}
\simeq{2\pi\over\mu_{\mu e}B_{\perp}}.\eqno(32)$$ It is clear that
this resonance could take place only when the value of
$2\Delta^{12}c_{2\omega}$ will be compensated by the magnetic
field twisting. Comparing the expression (31) with an analogous
one for the $\nu_{eL}\to\overline{\nu}_{\mu R}$ resonance we make
sure that they are mutually exclusive. Really, the fulfilment of
the $\nu_{eL}\to\overline{\nu}_{\mu R}$ resonance condition will
take place at negative values of $\dot{\Phi}$, while the $\nu_{\mu
L}\to\overline{\nu}_{eR}$ resonance condition demands positive
values of $\dot{\Phi}$.

From the obtained equations for the resonance conditions, the
oscillation lengths and the resonance widths we see that the
contributions coming from the AM and the CNR can be safely
neglected when the neutrino has the Majorana nature.

Further we proceed to the Dirac neutrino case. Here the electron
neutrinos could undergo three following resonance conversions
$$\nu_{eL}\to\nu_{\mu L},\qquad
\nu_{eL}\to{\nu}_{e R}, \qquad \nu_{eL}\to{\nu}_{\mu R}.$$ The
$\nu_{eL}\to\nu_{\mu L}$ resonance is of little interest. As in
the Majorana neutrino case it occurs before the convective zone.

The resonance condition and the maximal value of the oscillation
length for the $\nu_{eL}\to{\nu}_{e R}$ resonance are given by the
expressions
$$\Sigma_{\nu_{eL}{\nu}_{eR}}^D=V_{eL}+
{\cal{A}}^{DL}_{ee}-\dot{\Phi}=0.\eqno(33)$$
$$(L_{\nu_{eL}{\nu}_{eR}})_{max}
\simeq{2\pi\over\mu_{ee}B_{\perp}}.\eqno(34)$$ The situation, when
the term proportional to $[{\mbox{rot}}\ {\bf{H}}(z)]_z$ is
negligibly small compared to $\dot{\Phi}$ and the resonance
condition reduces to
$$V_{eL}\simeq\dot{\Phi},\eqno(35)$$ is not realistic.
Genuinely, in order to satisfy Eq. (33) it is necessary that the
twisting magnetic field exists over the distance being much bigger
than the solar radius. On the other hand, as we have already seen,
the quantity proportional to $[\mbox{rot}\ {\bf{H}}(z)]_z$ could
reach values of $10^{-30}$ eV and being negative it could
compensate the term of $V_{eL}$ in Eq.(33). In doing so the
$\nu_{eL}\to{\nu}_{e R}$ resonance may take place only in the
corona.

We are coming now to the $\nu_{eL}\to{\nu}_{\mu R}$ resonance. In
this case the pertinent expressions are as follows
$$\Sigma_{\nu_{eL}{\nu}_{\mu R}}^D=-2\Delta^{12}c_{2\theta}+V_{eL}+
{\cal{A}}_{ee}^{DL}-\dot{\Phi}=0,\eqno(36)$$
$$(L_{\nu_{eL}{\nu}_{\mu R}})_{max}\simeq{2\pi\over
\mu_{e\mu}B_{\perp}}.\eqno(37)$$ Having compared the foregoing
expressions with (24)-(26) one may come to the conclusion that the
conditions of observing the $\nu_{eL}\to{\nu}_{\mu R}$ resonance
are only slightly different from each other in both Dirac and
Majorana cases. Then, considering this resonance in the region of
the CS's we may argue, as in the Majorana case, that
$\nu_{eL}\to{\nu}_{\mu R}$ resonance may also occur only at the
cost of the magnetic field. The value of
$\Delta^{12}c_{2\theta}\simeq10^{-12}$ eV entering into the
resonance condition (36) could be compensated by the twisting
frequency $\dot{\Phi}$ only.

As for the $\nu_{\mu L}$ neutrinos produced in the MSW resonance,
they could undergo the following resonance conversions $\nu_{\mu
L}\to{\nu}_{eR}$ and $\nu_{\mu L}\to{\nu}_{\mu R}$. Their
resonance conditions coincide and will look like
$$\Sigma_{\nu_{\mu L}{\nu}_{eR}}^D=\Sigma_{\nu_{\mu L}
{\nu}_{\mu R}}^D=2\Delta^{12}c_{2\theta}+V_{\mu
L}+{\cal{A}}_{\mu\mu}^L -\dot{\Phi}=0.\eqno(38)$$ We see that the
obtained expressions have practically the same form as the
$\nu_{\mu L}\to\overline{\nu}_{eR}$ resonance condition.
Therefore, one may state that the $\nu_{\mu L}\to{\nu}_{eR}$ and
$\nu_{\mu L}\to{\nu}_{\mu R}$ resonances exhibit the identical
behavior with the $\nu_{\mu L}\to\overline{\nu}_{eR}$ resonance.
As a result, if the $\nu_{\mu L}\to{\nu}_{eR}$ and $\nu_{\mu
L}\to{\nu}_{\mu R}$ resonances are allowed then the
$\nu_{eL}\to\nu_{\mu R}$ resonance will be forbidden, and
conversely. From the foregoing equations follows that the AM and
the NCR should be taken into account for the Dirac neutrino case.

So, now we can write the expressions for the survival
probabilities of the electron neutrinos as follows
$${\cal{P}}^D_{\nu_{eL}\nu_{eL}}=1-({\cal{P}}_{\nu_{eL}{\nu}_{eR}}+
{\cal{P}}_{\nu_{eL}{\nu}_{\mu R}})\eqno(39)$$ for the Dirac
neutrino case, and
$${\cal{P}}^M_{\nu_{eL}\nu_{eL}}=1-{\cal{P}}_{\nu_{eL}\overline{\nu}_{\mu R}}
\eqno(30)$$ for the  Majorana neutrino case, where the
contribution of the MSW-resonance has been eliminated for reasons
expounded above.

Let us introduce the quantity which characterizes the weakening of
the electron neutrino beam. The weakening caused by the the
$\nu_{eL}\to\nu_{\mu R}$ resonance in the Dirac neutrino case has
the view
$$\eta_{\nu_{eL}{\nu}_{\mu R}}={N_{i}-N_{f}\over
N_{i}},$$ where $N_{i}$ and $N_{f}$ are numbers of the $\nu_{eL}$
neutrinos before and after the passage of the
$\nu_{eL}\to{\nu}_{\mu R}$ resonance, respectively. In order to
exactly estimate the value of $\eta_{\nu_{eL}\to{\nu}_{\mu R}}$
we should concretize the dependence on distance of the quantities
$n_e(z), n_n(z)$, $B_{\perp}(z)$ and solve Eq.(14). However, to
roughly estimate this quantity, it will suffice to compare the
resonance widths $\Gamma_{\nu_{eL}\nu_{\mu L}}$ and
$\Gamma_{\nu_{eL}{\nu}_{\mu R}}$, while taking into account the
value of $\eta_{\nu_{eL}\nu_{\mu L}}$. Calculations result in
$$\eta_{\nu_{eL}{\nu}_{\mu R}}\simeq\left\{\begin{array}{ll}
2\times10^{-4},\qquad\mbox{when}\
\mu_{e\mu}=(\mu_{ee})_{upper}=2.9\times10^{-11}\mu_B,\
\ B_{\perp}=10^5\ G,\\[2mm]
0.12,\hspace{19mm}\mbox{when}\
\mu_{e\mu}=(\mu_{\mu\mu})_{upper}=6.8\times10^{-10}\mu_B,\
B_{\perp}=10^7\ G.
\end{array}\right.\eqno(29)$$
It should be noted that all the magnetic-induced resonances have
the resonance widths which are completely determined by the
quantity $\mu_{\nu_l\nu_{l^{\prime}}}B_{\perp}$ and, as a result,
the foregoing estimations remain valid for such resonance
conversions. Hence, the total weakening of the electron neutrino
beam will be defined by the expression
$$\eta_{\nu_{eL}{\nu}_{\mu R}}+\eta_{\nu_{eL}{\nu}_{eR}}\simeq2
\eta_{\nu_{eL}{\nu}_{\mu R}}.\eqno()$$

So, the effect of weakening the electron neutrino beam because of
magnetic induced resonances, no matter how small it is, exists and
the problem is only in its registration with a help of detectors.
Let us explore whether second generation neutrino detectors can
observe this weakening. The work of these detectors are based on
coherent elastic (anti)neutrino-atomic nucleus scattering
(CE$\nu$NS). This type of low-energy (anti)neutrino interaction
was predicted in 1974\cite{KF74,FD74} and was recently discovered
by COHERENT Collaboration \cite{AK17}. It was shown that neutrinos
and antineutrinos of all types can elastically coherently interact
with all nucleons of the nucleus by means of a neutral current,
provided that the momentum transferred to the nucleus is small.
The cross section of such a process is relatively large, it is
more than two orders of magnitude (for heavy nuclei) larger than
the cross section of other known processes of interaction of
low-energy neutrinos. Such detectors are already being used for
monitoring the operation of a nuclear reactor in the on-line
regime. Examples are found in Russian Emission Detector-100
(RED-100) at Kalininskaya nuclear power plant \cite{AKIM}.
Installed at a distance of 19 meters from a nuclear reactor, where
the reactor antineutrino flux reaches the values
$1.35\times10^{13}\ \mbox{cm}^{-2}\ \mbox{c}^{-1}$, RED-100 should
record 3300 antineutrino events per day. Moreover, in the future,
it is planned to scale the detector by a factor of 10 to the mass
of the sensitive volume of the order of 1 ton
(RED-1000)\cite{AYA1}. This will make it possible to register
33,000 antineutrino events per day. Detectors designed to study
solar neutrinos will also operate on the analogous principle. Only
here we shall already talk about the coherent elastic
neutrino-nucleus scattering. Obviously, to assess the capabilities
of the second-generation solar neutrino detectors, one can use the
parameters of reactor neutrino detectors. For example, if RED-1000
will be used for detection of the solar $pp$-neutrinos, then it
could detect about 2000 neutrino events per day. Since in the
Dirac neutrino case the $\nu_{eR}$ and $\nu_{\mu R}$ are sterile
particles then the detectors utilizing the CE$\nu$NC will work in
the "disappearance" regime.

In the Majorana theory the right-handed neutrinos
$\overline{\nu}_{\mu R}$ and $\overline{\nu}_{eR}$ are physical
particles whereas detectors based on CE$\nu$NC are flavor-blind
(at least with the existing experimental technique). Then, since
the total neutrino flux is kept constant after traveling the
resonances, the detectors will not feel the change in the flavor
composition of the neutrino beam.

\section{Three-neutrino generations}
Let us consider the manner in which the inclusion of the third
neutrino generation will influence the oscillations picture. For
the Majorana neutrinos in the flavor basis the evolution equation
will look like
$$i{d\over dz}\left(\matrix{\nu_{eL}\cr\nu_{\mu
L}\cr\nu_{\tau L}\cr\overline{\nu}_{eR}\cr\overline{\nu}_{\mu
R}\cr\overline{\nu}_{\tau
R}}\right)=\Big({\cal{H}}_0^M+{\cal{H}}_{int}^M\Big)
\left(\matrix{\nu_{eL}\cr\nu_{\mu L}\cr\nu_{\tau
L}\cr\overline{\nu}_{eR}\cr\overline{\nu}_{\mu
R}\cr\overline{\nu}_{\tau R}}\right),\eqno(40)$$ where
$${\cal{H}}^M_0={\cal{U}}\left(\matrix{E_1&0&0&0& 0&0\cr
0&E_2&0&0&0&0\cr0&0&E_3&0 &0&0\cr 0&0&0&E_1&0&0\cr
0&0&0&0&E_2&0\cr0&0&0&0&0&E_3}\right){\cal{U}}^{-1}$$

$${\cal{H}}^M_{int}=$$
$$=\left(\matrix{V_{eL}^{\prime}+{\cal{A}}^L_{ee}&{\cal{A}}^L_{e\mu}
&{\cal{A}}^L_{e\tau}&0&\mu_{e\mu}B_{\perp}e^{-i\Phi}
&-\mu_{e\tau}B_{\perp}e^{-i\Phi}\cr {\cal{A}}^L_{\mu e}&V_{\mu L}+
{\cal{A}}^L_{\mu\mu}&{\cal{A}}^L_{\mu\tau}&-\mu_{e\mu}B_{\perp}e^{-i\Phi}
&0&\mu_{\mu\tau} B_{\perp}e^{-i\Phi}\cr{\cal{A}}^L_{\tau
e}&{\cal{A}}^L_{\tau\mu}&V_{\tau
L}+{\cal{A}}^L_{\tau\tau}&\mu_{e\tau}B_{\perp}e^{-i\Phi}&-\mu_{\mu\tau}
B_{\perp}e^{-i\Phi}&0&
\cr0&-\mu_{e\mu}B_{\perp}e^{i\Phi}&\mu_{e\tau}B_{\perp}e^{i\Phi}&-
V^{\prime}_{eL}+{\cal{A}}^R_{ee}
&{\cal{A}}^R_{e\mu}&{\cal{A}}^R_{e\tau}\cr
\mu_{e\mu}B_{\perp}e^{i\Phi}&0&-\mu_{\mu\tau}B_{\perp}e^{i\Phi}&{\cal{A}}^R_{\mu
e}&{-V_{\mu
L}+\cal{A}}^R_{\mu\mu}&{\cal{A}}^R_{\mu\tau}\cr-\mu_{e\tau}B_{\perp}e^{i\Phi}&
\mu_{\mu\tau} B_{\perp}e^{i\Phi}&0&{\cal{A}}^R_{\tau
e}&{\cal{A}}^R_{\tau\mu}&-V_{\tau
L}+{\cal{A}}^R_{\tau\tau}}\right),\qquad{\cal{U}}=\left(\matrix{\cal{D}&
0\cr 0&\cal{D}}\right),$$
$${\cal{U}}=\left(\matrix{\cal{D}& 0\cr 0&\cal{D}}\right),
\qquad{\cal{D}}=\exp({i\lambda_7\psi})\exp({i\lambda_5\phi})
\exp({i\lambda_2\omega}),$$ the $\lambda$'s are Gell-Mann matrices
corresponding to the spin-one matrices of the $SO(3)$ group,
$\psi=\theta_{23}, \ \phi=\theta_{13}, \ \omega=\theta_{12},$
$s_{\psi}=\sin\psi,\ c_{\psi}=\cos\psi,$ and so on. We remind that
the current values on the oscillation angles are \cite{FC17}
$$\sin^2\theta_{12}\simeq0.297,\qquad\sin^2\theta_{13}\simeq0.0215,
\qquad \sin^2\theta_{23}\simeq0.425.$$

Even though we work with the three component neutrino wave
function $\Psi^T=(\nu_{eL},\nu_{\mu L},\nu_{\tau L})$ the analysis
of the neutrino system behavior represents cumbersome process
\cite{VB89}. On the other hand, one could simplify the problem at
the cost of the change-over to a new basis. Let us demand that in
the new basis
$$\Psi^{\prime
M}=\left(\matrix{\nu_{1L}^{M}\cr\nu_{2L}^{M}\cr\nu_{3L}^{M}
\cr\overline{\nu}_{1R}^{M}\cr\overline{\nu}_{2R}^{M}\cr
\overline{\nu}_{3R}^{M}}\right),$$ which we call a "hatched"
basis, the Hamiltonian ${\cal{H}}_0^M$ will depend on the angle
$\omega$ only, while the Hamiltonian ${\cal{H}}^M_{int}$ depends
on the angles $\phi$ and $\psi$. In so doing, when in this basis
the angles $\psi$ and $\phi$ tend to zero, our results must be
converted into those obtained within the two flavor approximation
(FA). The hatched basis is connected with the flavor one in the
following manner
$$\Psi^{\prime M}={\cal{U}}^{\prime} \left(\matrix{\nu_{eL}\cr\nu_{\mu
L}\cr\nu_{\tau L}\cr\overline{\nu}_{eR}\cr\overline{\nu}_{\mu
R}\cr\overline{\nu}_{\tau R}}\right),\eqno(41)$$ where
$${\cal{U}}^{\prime}=\left(\matrix{{\cal{D}}^{\prime}& 0\cr
0&{\cal{D}}^{\prime}}\right),\qquad {\cal{D}}^{\prime}=
\exp({-i\lambda_5\phi})\exp({-i\lambda_7\psi})=
\left(\matrix{c_{\phi} &-s_{\phi}s_{\psi}&-s_{\phi}c_{\psi}\cr 0 &
c_{\psi}&-s_{\psi}\cr s_{\phi}& c_{\phi}s_{\psi}&
c_{\phi}c_{\psi}}\right).$$ Since the angle $\phi$ is much less
than the angles $\psi$ and $\omega$ then in the new basis the
$\nu^{M}_{1L}$ ($\overline{\nu}^{M}_{1R}$) state is predominantly
the $\nu_{eL}$ ($\overline{\nu}_{eR}$) one. Moreover, the
relations
$$\nu^{M}_{1L}\ \Big|_{\phi=0}=\nu_{eL},\qquad
\overline{\nu}^{M}_{1R}\
\Big|_{\phi=0}=\overline{\nu}_{eR}\eqno(42)$$ take place.

Inasmuch as the experimental bounds on the charge radiuses and the
anapole moments of all neutrino types are of the same order, for
the sake of simplicity we shall assume
$${\cal{A}}^{L,R}_{ee}={\cal{A}}^{L,R}_{\mu\mu}={\cal{A}}^{L,R}_{\tau\tau}=
{\cal{A}}^{L,R}_{ll},\qquad
{\cal{A}}^{L,R}_{e\mu}={\cal{A}}^{L,R}_{e\tau}={\cal{A}}^{L,R}_{\mu\tau}=
{\cal{A}}^{L,R}_{ll^{\prime}}.\eqno(43)$$ Then in the hatched
basis after the passage to the reference frame which rotates with
the same velocity as the magnetic field the Hamiltonians will look
like
$${\cal{H}}^{\prime M}_0=
\left(\matrix{-\Delta^{12}c_{2\omega} & \Delta^{12} s_{2\omega}&0
&0&0&0\cr \Delta^{12}s_{2\omega}&\Delta^{12}c_{2\omega}&0&0&0&0\cr
0&0&\Delta^{31}+\Delta^{32}&0&0&0\cr 0&0&0&
-\Delta^{12}c_{2\omega} & \Delta^{12} s_{2\omega}&0\cr
0&0&0&\Delta^{12}s_{2\omega}&\Delta^{12}c_{2\omega}&0\cr
0&0&0&0&0&\Delta^{31}+\Delta^{32}}\right),\eqno(44)$$
$${\cal{H}}^{\prime M}_{int}=\left(\matrix{\Lambda_{11}-\dot{\Phi}/2&
\Lambda_{12} &\Lambda_{13}& 0
&\mu_{12}B_{\perp}&\mu_{13}B_{\perp}\cr \Lambda_{12} &
\Lambda_{22}-\dot{\Phi}/2
&\Lambda_{23}&-\mu_{12}B_{\perp}&0&-\mu_{23}B_{\perp}\cr
\Lambda_{13}& \Lambda_{23}&
\Lambda_{33}-\dot{\Phi}/2&-\mu_{13}B_{\perp}&\mu_{23}B_{\perp}&0\cr
0&-\mu_{12}B_{\perp}&-\mu_{13}B_{\perp}&
\overline{\Lambda}_{11}+\dot{\Phi}/2& \overline{\Lambda}_{12}
&\overline{\Lambda}_{13}\cr \mu_{12}B_{\perp}& 0 &
\mu_{23}B_{\perp}&
\overline{\Lambda}_{12}&\overline{\Lambda}_{22}+\dot{\Phi}/2&
\overline{\Lambda}_{23}\cr\mu_{13}B_{\perp}&-\mu_{23}B_{\perp}&0
&\overline{\Lambda}_{13} &\overline{\Lambda}_{23}&
\overline{\Lambda}_{33}+\dot{\Phi}/2}\right),\eqno(45)$$ where
$$\mu_{12}=\mu_{e\mu}c_{\psi}c_{\phi}+\mu_{e\tau}s_{\psi}c_{\phi}+
\mu_{\mu\tau}s_{\phi},\qquad \mu_{13}=
\mu_{e\mu}s_{\psi}-\mu_{e\tau}c_{\psi},$$
$$\mu_{23}=
-\mu_{e\mu}c_{\psi}s_{\phi}-\mu_{e\tau}s_{\psi}s_{\phi}+\mu_{\mu\tau}c_{\phi},$$
$$\Lambda_{11}=(V_{eL}^{\prime}-V_{\mu L})c_{\Phi}^2+V_{\mu L}+
{\cal{A}}^{L}_{ll}-2{\cal{A}}^{L}_{ll^{\prime}}[c_{\Phi}s_{\Phi}(c_{\psi}+s_{\psi})-s_{\Phi}^2
c_{\psi}s_{\psi}],$$
$$\Lambda_{22}=V_{\mu L}+{\cal{A}}^{L}_{ll}-
2{\cal{A}}^{L}_{ll^{\prime}}s_{\psi}c_{\psi},$$
$$\Lambda_{33}=V_{\mu L}+{\cal{A}}^{L}_{ll}+
2{\cal{A}}^{L}_{ll^{\prime}}[c_{\Phi}^2s_{\psi}c_{\psi}+
c_{\Phi}s_{\Phi}(c_{\psi}+s_{\psi})],$$
$$\Lambda_{12}=\Lambda_{21}=
{\cal{A}}^{L}_{ll^{\prime}}(c_{\psi}-s_{\psi})[c_{\Phi}+
s_{\Phi}(c_{\psi}+s_{\psi})],$$
$$\Lambda_{13}=\Lambda_{31}=(V_{eL}^{\prime}-V_{\mu
L})c_{\Phi}s_{\Phi}-2{\cal{A}}^{L}_{ll^{\prime}}[c_{\Phi}s_{\Phi}c_{\psi}-
(c_{\psi}+s_{\psi})(c_{\Phi}^2-s^2_{\Phi})],$$
$$\Lambda_{23}=\Lambda_{32}=
{\cal{A}}^{L}_{ll^{\prime}}(c_{\psi}-s_{\psi})[s_{\Phi}+
c_{\Phi}(c_{\psi}+s_{\psi})],$$
$$\overline{\Lambda}_{ik}=\Lambda_{ik}\Big(V_{eL}^{\prime}\to-V_{eL}^{\prime},
V_{\mu L}\to-V_{\mu L},
{\cal{A}}^{L}_{ll^{\prime}}\to{\cal{A}}^{R}_{ll^{\prime}},
{\cal{A}}^{L}_{ll}\to{\cal{A}}^{R}_{ll}\Big),\qquad i,k=1,2,3.
 $$

For the Dirac neutrinos in the basis $\Psi^{\prime D
T}=(\nu^D_{1L},\nu^{D}_{2L},\nu^{D}_{3L},\nu^D_{1R},
\nu^{D}_{2R},\nu^{D}_{3R})$ the free Hamiltonian
${\cal{H}}_0^{\prime D}$ coincides with ${\cal{H}}_0^{\prime M}$
while the interaction Hamiltonian takes the form
$${\cal{H}}^{\prime D}_{int}=\left(\matrix{\Lambda^D_{11}-\dot{\Phi}/2&
\Lambda_{12}^D &\Lambda_{13}^D& \mu_{ee}^{\prime}B_{\perp} &
\mu_{e\mu}^{\prime}B_{\perp}& \mu_{e\tau}^{\prime}B_{\perp}\cr
\Lambda_{12}^D & \Lambda_{22}^D-\dot{\Phi}/2 &\Lambda_{23}^D&
\mu_{\mu e}^{\prime}B_{\perp}& \mu_{\mu\mu}^{\prime}B_{\perp}&
\mu_{\mu\tau}^{\prime}B_{\perp}\cr \Lambda_{13}^D& \Lambda_{23}^D&
\Lambda_{33}^D-\dot{\Phi}/2& \mu_{\tau e}^{\prime}B_{\perp}&
\mu_{\tau\mu}^{\prime}B_{\perp}&
\mu_{\tau\tau}^{\prime}B_{\perp}\cr
\mu_{ee}^{\prime}B_{\perp}&\mu_{e\mu}^{\prime}B_{\perp}&\mu_{e\tau}^{\prime}
B_{\perp}&\dot{\Phi}/2& 0 &0\cr \mu_{\mu e}^{\prime}B_{\perp}&
\mu_{\mu\mu}^{\prime}B_{\perp} & \mu_{\mu\tau}^{\prime}B_{\perp}&
0&\dot{\Phi}/2& 0\cr \mu_{\tau
e}^{\prime}B_{\perp}&\mu_{\tau\mu}^{\prime}B_{\perp}&\mu_{\tau\tau}^{\prime}B_{\perp}
&0&0& \dot{\Phi}/2}\right),\eqno(46)$$ where
$$\Lambda_{ik}^D=\Lambda_{ik}(V^{\tilde{\delta}}_{ee}\to 0,
{\cal{A}}^L_{ll^{\prime}}\to{\cal{A}}^{DL}_ {ll^{\prime}}),$$
$$\left(\matrix{\mu^{\prime}_{ee}B_{\perp}&\mu^{\prime}_{e\mu}
B_{\perp}&\mu^{\prime}_{e\tau} B_{\perp}\cr
\mu^{\prime}_{e\mu}B_{\perp}& \mu^{\prime}_{\mu\mu}B_{\perp}&
\mu^{\prime}_{\mu\tau}B_{\perp}\cr
\mu^{\prime}_{e\tau}B_{\perp}&\mu^{\prime}_{\mu\tau}B_{\perp}&\mu^{\prime}
_{\tau\tau}B_{\perp}}\right)=
{\cal{D}}^{\prime}\left(\matrix{\mu_{ee}B_{\perp}&\mu_{e\mu}
B_{\perp}&\mu_{e\tau} B_{\perp}\cr \mu_{e\mu}B_{\perp}&
\mu_{\mu\mu}B_{\perp}& \mu_{\mu\tau}B_{\perp}\cr
\mu_{e\tau}B_{\perp}&\mu_{\mu\tau}B_{\perp}&\mu_{\tau\tau}B_{\perp}}\right)
{\cal{D}}^{\prime -1}.$$ Since in the Dirac neutrino case the
masses of all singly charged Higgs bosons lay at the TeV scale,
then in the expression for ${\cal{H}}^{\prime D}_{int}$ we have
neglected their contributions. We should also focus our attention
on the fact that in the solar atmosphere all elements of
${\cal{H}}_{int}^{\prime M,D}$ are much more less than the ones of
${\cal{H}}_0^{\prime M,D}$. So, the perturbation theory may be
applied.

Now we proceed to the investigation of the resonance transitions
in the neutrino system under study. Assuming the Majorana neutrino
nature we start our discussion from the $\nu_{1L}^{M
}\to\nu_{2L}^{M}$ transition. The resonance condition and the
maximal value of the oscillation length are as follows
$$\Sigma_{\nu_{1L}\nu_{2L}}=-2\Delta^{12}c_{2\omega}+
(V_{eL}^{\prime}-V_{\mu L})c_{\Phi}^2-
2{\cal{A}}^{L}_{ll^{\prime}}[c_{\Phi}s_{\Phi}(c_{\psi}+s_{\psi})-(1+s_{\Phi}^2)
c_{\psi}s_{\psi}]=0,\eqno(47)$$
$$(L_{\nu_{1L}^{M}\nu^{M}_{2L}})_{max}=
{2\pi\over\Delta^{12}s_{2\omega}+
{\cal{A}}^{L}_{ll^{\prime}}(c_{\psi}-s_{\psi})[c_{\phi}+
s_{\phi}(c_{\psi}+s_{\psi})]}.\eqno(48)$$ Corresponding
expressions for the $\nu_{1L}^{D}\to\nu_{2L}^{D}$ transition
follow from Eqs.(47),(48) under the replacement
$V^{\tilde{\delta}}_{ee}\to 0$ and
${\cal{A}}^L_{ll^{\prime}}\to{\cal{A}}^{DL}_ {ll^{\prime}}$. When
$\psi=\phi=0$ the expressions (47) and (48) convert to the
resonance condition and the oscillation length for the MSW
resonance in two FA (recall that we have set
${\cal{A}}_{ee}^L={\cal{A}}^L_{\mu\mu}$). That allows us to
believe the $\nu_{1L}^{M }\to\nu_{2L}^{M}$-resonance as an analog
of the $\nu_{eL}^{M }\to\nu_{\mu L}^{M}$ resonance in the two FA.
Moreover, by virtue of the fact
$${|\Sigma_{\nu_{1L}\nu_{2L}}-\Sigma_{\nu_{eL}\nu_{\mu L}}|
\over2\Delta^{12}c_{2\omega}}\ll1,$$ both resonances are
characterized by the identical formulas. However, for the reasons
stated above, this resonance is of no interest for us and we pass
to considering the $\nu_{1L}^{M}\to\nu_{3L}^{M}$ and $\nu_{1L}^{
D}\to\nu_{3L}^{D}$ resonances. In the Hamiltonians
${\cal{H}}^{\prime M}$ and ${\cal{H}}^{\prime D}$ the quantity
$\Sigma=\Delta^{31}+\Delta^{32}$ is present. Since it offers the
dominant term then the $\nu_{3L}^{M}$ and $\nu_{3L}^{D}$ states
are decoupled from the remaining ones (except the
${\overline{\nu}}_{3R}^{M}$ and $\nu_{3R}^{D}$ states). As a
result the $\nu_{1L}^{M}\to\nu_{3L}^{M}$ and
$\nu_{1L}^{D}\to\nu_{3L}^{D}$ oscillations controlled by the
$\Sigma$-term could be simply averaged out in the final survival
probability for neutrinos of any flavor.

Further we pass to discussion of the magnetic-induced resonances
which could take place in the regions of the CS's. Let us begin
with the $\nu_{1L}^{M}\to\overline{\nu}_{3R}^{M}$ and
$\nu_{1L}^{D}\to\nu_{3R}^{D}$ resonances. They are also controlled
by the $\Sigma$-term and, as a result, these resonances appear to
be forbidden.

The $\nu_{1L}^{M}\to\overline{\nu}_{1R}^{M}$ and
$\nu_{1L}^{D}\to\nu_{1R}^{D}$ resonances are the next subject of
our investigation. In the Majorana neutrino case the resonance
width is equal to zero and, as a result, the
$\nu_{1L}^{M}\to\overline{\nu}_{1R}^{M}$ resonance is not
observed. For the Dirac neutrino the resonance condition and the
maximum value of the oscillation length are defined by the
following expressions
$$(V_{eL}-V_{\mu L})c_{\Phi}^2+V_{\mu L}+
{\cal{A}}^{DL}_{ll}-
2{\cal{A}}^{DL}_{ll^{\prime}}[c_{\Phi}s_{\Phi}(c_{\psi}+s_{\psi})-s_{\Phi}^2
c_{\psi}s_{\psi}]-\dot{\Phi}=0,\eqno(49)$$
$$(L_{\nu_{1L}^{D}\nu_{1R}^{D}})_{max}
\simeq{2\pi\over\mu_{ee}^{\prime}B_{\perp}}.\eqno(50)$$ When
$\phi=0$ the obtained expressions convert into the corresponding
ones for the $\nu_{eL}^D\to\nu_{eR}^D$ in two FA (see Eqs.(33) and
(34)). That allows us to consider the
$\nu_{1L}^{D}\to\nu_{1R}^{D}$ resonance as an analog of the
$\nu_{eL}^D\to\nu_{eR}^D$ resonance in two FA. Comparing the
resonance condition (49) with the analogous expression (33)
obtained in two FA, we see that they differ from one another by
the quantity being proportional to $\sin\phi$. Therefore, when the
condition (33) is fulfilled, then the same is true for the
condition (49). So, the $\nu_{1L}^D\to\nu_{1R}^D$ resonance may be
in existence in the solar corona.

In what follows we shall deal with the
$\nu_{1L}^{M}\to\overline{\nu}_{2R}^{M}$ and
$\nu_{1L}^{D}\to\nu_{2R}^{D}$ resonances. For the former the
resonance condition and the maximum value of the oscillation
length are as follows
$$-2\Delta^{12}c_{2\omega}+(V_{eL}^{\prime}-V_{\mu
L})c_{\Phi}^2+2V_{\mu L}+{\cal{A}}^{L}_{ll}-{\cal{A}}^R_{ll}+
2{\cal{A}}^R_{ll^{\prime}}s_{\psi}c_{\psi}-2{\cal{A}}^L_{ll^{\prime}}
[c_{\Phi}s_{\Phi}
(c_{\psi}+s_{\psi})-s_{\Phi}^2c_{\psi}s_{\psi}]-\dot{\Phi}=0,\eqno(51)$$
$$(L_{\nu_{1L}^{M}\overline{\nu}_{1R}^{M}})_{max}\simeq{2\pi\over
\mu_{12}B_{\perp}},\eqno(52)$$ while for the latter the
corresponding expressions will look like
$$-2\Delta^{12}c_{2\omega}+(V_{eL}-V_{\mu
L})c_{\Phi}^2+V_{\mu
L}+{\cal{A}}^{DL}_{ll}-2{\cal{A}}^{DL}_{ll^{\prime}}[c_{\Phi}s_{\Phi}
(c_{\psi}+s_{\psi})-s_{\Phi}^2c_{\psi}s_{\psi}]-\dot{\Phi}=0,\eqno(53)$$
$$(L_{\nu_1^{0D}\overline{\nu}_2^{D}})_{max}\simeq{2\pi\over
\mu_{e\mu}^{\prime}B_{\perp}}.\eqno(54)$$ Since at $\phi=\psi=0$
the foregoing expressions convert to the corresponding ones
obtained in two FA one may conclude that the investigated
resonances should be considered as the analogs of the
$\nu_{eL}^M\to\overline{\nu}_{\mu R}^M$ and
$\nu_{eL}^D\to\nu^D_{\mu R}$ resonances. The resonance condition
(51) differers from the analogous one arrived in two FA on the
quantities which are proportional to ${\cal{A}}_{ll^{\prime}}^L$
and ${\cal{A}}_{ll^{\prime}}^R$. These quantities are so much less
compared with others that we could neglect them. Therefore, the
resonance condition (51) is reduced to Eq.(24). As for the
expression (53), that is reduced to the analogous expression of
the two FA (36) when $\phi=0$. So, the
$\nu_{1L}^{M}\to\overline{\nu}_{2R}^{M}$ and $\nu_{1L}^{
D}\to\nu_{2R}^{D}$ resonances may also occur only at the cost of
magnetic field twisting only.

Further we also consider all possible resonance transitions of the
$\nu_{2L}^{M}$ and $\nu_{2L}^{D}$ states. The resonance condition
and the maximal oscillation length for the
$\nu_{2L}^{M}\to\overline{\nu}_{1R}^{M}$ transition will look like
$$2\Delta^{12}c_{2\omega}+(V_{eL}^{\prime}-V_{\mu L})c_{\Phi}^2+2V_{\mu L}
+{\cal{A}}^{L}_{ll}-{\cal{A}}^{R}_{ll}-2{\cal{A}}^{L}_{ll^{\prime}}s_{\psi}c_{\psi}+
2{\cal{A}}^{R}_{ll^{\prime}}[c_{\Phi}s_{\Phi}(c_{\psi}
+s_{\psi})-s_{\Phi}^2c_{\psi}s_{\psi}]-\dot{\Phi}=0,\eqno(55)$$
$$(L_{\nu_{2L}^{M}\overline{\nu}_{1R}^{M}})_{max}\simeq{2\pi\over
\mu_{12}B_{\perp}}.\eqno(56)$$ The corresponding expressions for
the $\nu_{2L}^{D}\to\nu_{1R}^{D}$ resonance are determined by the
following way
$$2\Delta^{12}c_{2\omega}+V_{\mu L}+{\cal{A}}^{DL}_{ll}-
2{\cal{A}}^{DL}_{ll^{\prime}}s_{\psi}c_{\psi}-\dot{\Phi}=0,\eqno(57)$$
$$(L_{\nu_{2L}^{D}\nu_{1R}^{D}})_{max}\simeq{2\pi\over
\mu_{e\mu}^{\prime}B_{\perp}}.\eqno(58)$$ Setting $\psi=\phi=0$ in
the expressions obtained we arrive at the resonance conditions and
the maximal oscillation lengths for the $\nu_{\mu
L}^{M}\to\overline{\nu}_{eR}^{M}$ and $\nu_{\mu
L}^{D}\to{\nu}_{eR}^{D}$ transitions, what permits us to consider
these resonances as the analogs of the $\nu_{\mu
L}^{M}\to\overline{\nu}_{eR}^{M}$ and $\nu_{\mu
L}^{D}\to\nu_{eR}^{D}$ in the two FA. It is clear that the
fulfilment (55) and (57) may take place only at
$$2\Delta^{12}c_{2\omega}\simeq\dot{\Phi}.\eqno(59)$$

Now we proceed to the treatment of the
$\nu_{2L}^{M}\to\overline{\nu}_{2R}^{M}$ and
$\nu_{2L}^{D}\to\nu_{2R}^{D}$ transitions. As far as the
$\nu_{2L}^{M}\to\overline{\nu}_{2R}^{M}$ transition is concerned,
that appears to be forbidden. The resonance condition and the
maximal oscillation length for the $\nu_{2L}^{D}\to\nu_{2R}^{D}$
transition are defined by the expressions
$$V_{\mu L}+{\cal{A}}^{DL}_{ll}-
2{\cal{A}}^{DL}_{ll^{\prime}}s_{\psi}c_{\psi}-\dot{\Phi}=0,\eqno(60)$$
$$(L_{\nu_{2L}^{D}\nu_{2R}^{D}})_{max}\simeq{2\pi\over
\mu_{\mu\mu}^{\prime}B_{\perp}}.\eqno(61)$$ When $\psi=\phi=0$ the
obtained formulae convert into the resonance condition and the
maximal oscillation length for the $\nu_{\mu L}^D\to\nu_{\mu R}^D$
resonance, that is, $\nu_{2L}^{D}\to\nu_{2R}^{D}$ resonance is the
analog of $\nu_{\mu L}^D\to\nu_{\mu R}^D$ resonance in two FA. The
fulfilment of (60) may be provided only when
$$\dot{\Phi}\approx0,\qquad V_{\mu L}\simeq-{\cal{A}}^{DL}_{ll}+
2{\cal{A}}^{DL}_{ll^{\prime}}s_{\psi}c_{\psi}\eqno(62)$$ that is
real in the Sun's conditions.

The contribution to the considered oscillation picture from the
third neutrino generation may come only from the
$\nu_{3L}^{D}\to\nu_{3R}^{D}$ resonance. The expressions being
pertinent to this resonance will look like
$$V_{\mu L}+{\cal{A}}^{DL}_{ll}+
2{\cal{A}}^{DL}_{ll^{\prime}}[c_{\Phi}^2s_{\psi}c_{\psi}+
c_{\Phi}s_{\Phi}(c_{\psi}+s_{\psi})]-\dot{\Phi}=0,\eqno(63)$$
$$(L_{\nu_{3L}^{D}\nu_{3R}^{D}})_{max}\simeq{2\pi\over
\mu_{\tau\tau}^{\prime}B_{\perp}}.\eqno(64)$$ It is clear that in
the resonance condition the matter potential $V_{\mu L}$ may be
compensated only by the terms which are proportional to
$\mbox{rot}\ {\bf{B}}.$

Again we see that the influence of the AM and the NCR on the
oscillation picture appears to be significant for the Dirac
neutrino case only.

In the neutrino physics, quantities which are measured in
experiments should be represented in the flavor basis. Therefore,
in the expressions for the resonance transition probabilities, we
should pass from the hatched basis to the flavor one. Let us
assume that solving the evolution equation both for the Majorana
and for the Dirac neutrinos we have determined all the transition
probabilities ${\cal{P}}^{M,D}(\nu_{iL}\to\nu_{kR})$
($i,k=1,2,3$). Then, taking into consideration the flavor content
of the $\psi^{\prime M,D}$ states we could find the probabilities
of the transitions between any flavor states. For example, the
$\nu^D_{\mu L}\to\nu^D_{\tau R}$ transition probability is as
follows
$${\cal{P}}^D(\nu_{\mu L}\to\nu_{\tau R})=
s_{\phi}^2\Big[s_{\phi}^2c_{\psi}^2s_{\psi}^2{\cal{P}}^D(\nu_{1L}\to\nu_{1R})+
s_{\psi}^4{\cal{P}}^D(\nu_{1L}\to\nu_{2R})+
c_{\psi}^4{\cal{P}}^D(\nu_{2L}\to\nu_{1R})\Big]+$$
$$+c_{\psi}^2s_{\psi}^2\Big[{\cal{P}}^D(\nu_{2L}\to\nu_{2R})+
c_{\phi}^4{\cal{P}}^D(\nu_{3L}\to\nu_{3R})\Big].\eqno(65)$$ As for
the electron neutrino survival probabilities are concerned, they
are given by the expressions
$${\cal{P}}^M({\nu_{eL}\to\nu_{eL}})=1-
[{\cal{P}}^M(\nu_{eL}\to\overline{\nu}_{e
R})+{\cal{P}}^M(\nu_{eL}\to\overline{\nu}_{\mu
R})+{\cal{P}}^M(\nu_{eL}\to\overline{\nu}_{\tau R})]+$$
$$=1-c_{\phi}^2{\cal{P}}^M({\nu_{1L}\to\overline{\nu}
_{2R}})\eqno(66)$$ in the Majorana neutrino case. and
$${\cal{P}}^D({\nu_{eL}\to\nu_{eL}})=1-\{c_{\phi}^2[{\cal{P}}^D({\nu_{1L}\to\nu_{1R}})+
{\cal{P}}^D({\nu_{1L}\to\nu_{2R}})]+s_{\phi}^2{\cal{P}}^D({\nu_{3L}
\to\nu_{3R}}\}\eqno(67)$$ in the Dirac neutrino case. It is easy
to check that when $\psi=\phi=0$ the expressions (66) and (67)
convert into the corresponding ones obtained in the two FA while
${\cal{P}}^D(\nu_{\mu L}\to\nu_{\tau R})$ becomes equal to zero.

\section{Conclusions}

The behavior of neutrinos, endowed by such multipole moments as
the charge radius, the magnetic and anapole moments, in intensive
magnetic fields has been explored within the left-right symmetric
model. It was assumed that the magnetic fields are vortex,
nonhomogeneous and have twisting nature. For the geometrical phase
$\Phi(z)=\arctan(B_y/B_x)$ connected with magnetic field twisting
$\dot{\Phi}(z)$ the simple model $\Phi=\exp[{\alpha z/L_{mf}]}$
($L_{mf}$ is the distance at which the magnetic field exists) has
been used. As the examples of such magnetic fields we have covered
fields of the coupled sunspots (CS's) being the sources of the
future solar flares. The investigations have been carried out both
for the Majorana and for the Dirac neutrinos. In the first stage
we have discussed the behavior of the neutrino beam in two flavor
approximation (FA). The evolution equation has been written in the
Schrodinger-like form and all the possible magnetic-induced
resonance conversions have been found. Further the problem has
been investigated in three FA. In order to lighten the analysis of
the resonance conversions and make the results physically more
transparent we have passed from the flavor basis to the new one
(hatched basis). In the new basis the free Hamiltonian
${\cal{H}}_0$ depends on the $\theta_{12}$ angle while the
interaction Hamiltonian ${\cal{H}}_{int}$ depends on the
$\theta_{23}$ and $\theta_{13}$ angles. The resonance conditions,
the transition widths and the oscillation lengths of all
magnetic-induced resonances have been found. The obtained
expressions are distinguished from the corresponding ones obtained
in the two FA only slightly. This situation is caused by choosing
the hatched basis in such a way that one state is predominantly
the $\nu_{eL}$-state while the rest of two are mixings of
$\nu_{\mu L}$- and $\nu_{\tau L}$-states. Taking into account the
flavor content of the hatched states we have expressed the
electron neutrino survival probability in terms of the
probabilities of the transitions between hatched states.

Under description of the neutrino oscillations the NCR
phenomenology is analogous to that of the AM. In the Majorana
neutrino case only nondiagonal elements of the NCR are different
from zero while the AM has both the nonzero diagonal and nonzero
nondiagonal elements. However in the Sun conditions these MM's do
not exert a marked influence on the values of the oscillation
parameters. On the other hand, when the neutrinos have the Dirac
nature the nonzero diagonal elements of the NCR and the AM could
lead to the appearance of new resonances. Using the upper bounds
on the NCR
$$|<r^2>|=few\times10^{-32}\ \mbox{cm}^2$$
and the values of the current producing the CS's magnetic field
$$j=10^{-1}\ \mbox{A}/\mbox{cm}^2,$$ one may get for the contribution
connected with these MM's the value of the same order as the
corona matter potential $\sim10^{-30}$ eV. Therefore, the
resonances initiated by the AM and the NCR may take place in the
solar corona. In so doing introducing the NCR changes the
resonance position and in specific cases could cause the vanishing
of the resonance.

For all the magnetic-induced resonances the oscillation width
depends on the quantity $\mu_{ll^{\prime}}B_{\perp}$ which, in its
turn, determines the weakening of the electron neutrino beams
$\eta_{\nu_{eL}\nu_{xR}}$. For example, when
$\mu_{ll^{\prime}}=6.8\times10^{-10}\mu_B$ and $B_{\perp}=10^8$ G
we have $\eta_{\nu_{eL}\nu_{xR}}\simeq1.2$. So in the case of the
super solar flares we have a good chance to detect the weakening
of the electron neutrino beam caused by the resonance conversions
$\nu_{eL}\to\overline{\nu}_{\mu R}$ and
$\nu_{eL}\to\overline{\nu}_{\tau R}$ (Majorana neutrino case) or
$\nu_{eL}\to\nu_{eR}$, $\nu_{eL}\to\nu_{\mu R}$ and
$\nu_{eL}\to\nu_{\tau R}$ (Dirac neutrino case). It should be
stressed that in the Dirac neutrino case all magnetic-induced
resonances transfer active neutrinos into sterile ones while in
the Majorana neutrino case we deal with active neutrinos only.
Decreasing of the electron neutrino flux which passes through the
magnetic field region during the initial solar flare stage could
be detected at the neutrino detectors of the next generation whose
work is based on the reaction of the coherent elastic
neutrino-nucleus scattering.

It should be stressed that the flares could take place in Sun-like
stars as well. In that case the super-flares present a severe
hazard to astronauts. Therefore, the problem of the flare
forecasting is actual for the cosmic flights as well. Obviously
that terrestrial neutrino detectors will be useless when flying
outside the solar system. The problem can be solved with the help
of a detector similar in design to the RED-100 installed on a
spacecraft. This detector can operate in the mode on
"disappearance" of electron neutrinos with a certain wavelength.

It might be worth pointing out the connection between our results
and the observations of decreasing the $\beta$-decay rates of some
elements during the initial stage of the solar flare
\cite{JHJ2009,DOK13}. According to Refs. \cite{TM16,PA18} this
phenomena is caused by the depletion of the solar electron
neutrinos (the hypothesis of the $\nu_{eL}$-induced $\beta$
decays). Then one may state that decreasing the $\beta$-decay
rates is the experimental confirmation of the resonance
conversions of the $\nu_{eL}$ neutrinos when they pass through the
CS's magnetic fields.

\section*{Acknowledgments}
This work is partially supported by the grant of Belorussian
Ministry of Education No 20211660

\end{document}